\documentclass[a4paper,12pt]{article}

\usepackage[margin=1in,dvips,a4paper,nohead]{geometry}
\usepackage[margin=1em, font=small, justification=centerlast,
tableposition=top]{caption}

\usepackage{amsfonts,amssymb,amsmath}
\usepackage[noBBpl,slantedGreek]{mathpazo}

\usepackage{pstricks}

\psset{unit=1pt, dimen=middle, linewidth=.5pt, arrowsize=3pt 2}

\newdimen \psx
\newdimen \psy

\def\psddots (#1,#2){
  \psx #1\psxunit
  \psy #2\psyunit
  \qdisk(\psx,\psy){.7pt}
  \advance \psx -.2\psxunit
  \advance \psy .2\psyunit
  \qdisk(\psx,\psy){.7pt}
  \advance \psx .4\psxunit
  \advance \psy -.4\psyunit
  \qdisk(\psx,\psy){.7pt}
}

\def\pssddots (#1,#2){
  \psx #1\psxunit
  \psy #2\psyunit
  \qdisk(\psx,\psy){.7pt}
  \advance \psx .2\psxunit
  \advance \psy .2\psyunit
  \qdisk(\psx,\psy){.7pt}
  \advance \psx -.4\psxunit
  \advance \psy -.4\psyunit
  \qdisk(\psx,\psy){.7pt}
}

\makeatletter
\def\numberbysection{\@addtoreset{equation}{section}
        \def\theequation{\arabic{equation}}}

\def \:{\mskip .5\thinmuskip}
\def\ph {{\hbox to 0pt{\phantom{$\scriptstyle -1$}\hss}}}

\def \id{\mathop{\mathrm{id}}\nolimits}

\def\bbC {\mathbb C}

\def\bbR {\mathbb R}
\def\bbZ {\mathbb Z}

\def\calL {\mathcal L}
\def\calM {\mathcal M}
\def\calN {\mathcal N}

\mathchardef\Gamma "100

\def\gothgl{\mathfrak{gl}}

\def\rme {\mathrm e}

\def\rmi {\mathrm i}
\def\rmGL {\mathrm{GL}}

\def\wt {\widetilde}
\def\Mat {\mathrm{Mat}}

\title{\bf More non-Abelian loop Toda solitons}

\author{Kh. S. Nirov\thanks{On leave of absence from the
\small \em Institute for Nuclear Research of the Russian Academy of
Sciences, 60th October Ave 7a, 117312 Moscow,
Russia}~~ and~ A. V. Razumov\thanks{On leave of absence from the
\small \em Institute for High Energy Physics, 142281 Protvino,
Moscow Region, Russia}\\
\small Fachbereich C--Physik, Bergische Universit\"at Wuppertal\\
\small D-42097 Wuppertal, Germany
}

\date{}

\begin{document}

\maketitle

\begin{abstract}
We find new solutions, including soliton-like ones, for a
special case of non-Abelian loop Toda equations associated
with complex general linear groups. We use the method of
rational dressing based on an appropriate block-matrix
representation suggested by the $\bbZ$-gradation under
consideration. We present solutions in a form of a direct
matrix generalization of the Hirota's soliton solution
already well-known for the case of Abelian loop Toda
systems.
\end{abstract}

\begin{center}
\small
{{\bf Mathematics Subject Classification (2000).} 37K10, 37K15, 35Q58}
\\
{{\bf Keywords.} Non-Abelian loop Toda systems, rational dressing
method, \\ soliton-like solutions}
\end{center}

\section{Introduction}

The two-dimensional Toda equations play an essential r\^ole in
understanding certain structures in classical and quantum integrable
systems. They are formulated as nonlinear partial differential
equations of second order and are associated with Lie groups, see,
for example, the monographs \cite{LezSav92, RazSav97}. The Toda
equations associated with affine Kac--Moody groups are of special
interest, because they possess soliton solutions having a lot of
physical applications. The simplest example here is the celebrated
sine-Gordon equation known for a long time. Another example of an
affine Toda equation was constructed in paper \cite{Mik79} as a
direct two-dimensional generalization of the famous mechanical Toda
chain. Later on, the consideration of paper \cite{Mik79} was
generalized in papers \cite{Mik81, MikOlsPer81} in the case of Toda
chains related to various affine Kac--Moody algebras. Another
approach to formulating affine Toda systems, based on folding
properties of Dynkin diagrams, was implemented in \cite{KhaSas96}.
Note also that papers \cite{Mik79, Mik81, MikOlsPer81, OliTurUnd86}
were pioneering in investigating the question of integrability of
affine Toda field theories by considering the corresponding
zero-curvature representation.

It is convenient to consider instead of the Toda systems associated
with affine Kac-Moody groups the Toda systems associated with loop
groups. There are two reasons to do so. First, the affine Kac-Moody
groups can be considered as a loop extension of loop groups and the
solutions of the corresponding Toda equations are connected in a
simple way, see, for example, paper \cite{ConFerGomZim93}. Second,
in distinction to the loop groups, there is no a realization of
affine Kac--Moody groups suitable for practical usage.

It is known that a Toda equation associated with a Lie group is
specified by the choice of a $\bbZ$-gradation of its Lie algebra
\cite{LezSav92, RazSav97}. Hence, to classify the Toda systems
associated with some class of Lie groups one needs to describe all
$\bbZ$-gradations of the respective Lie algebras. Recently, in a
series of papers \cite{NirRaz06, NirRaz07a, NirRaz07b}, we
classified a wide class of Toda equations associated with untwisted
and twisted loop groups of complex classical Lie groups. More
concretely, we introduced the notion of an integrable
$\bbZ$-gradation of a loop Lie algebra, and found all such
gradations with finite-dimensional grading subspaces for the loop
Lie algebras of complex classical Lie algebras. Then we described
the respective Toda equations. It appeared that despite the fact
that we consider Toda equations associated with infinite-dimensional
Lie groups, the resulting Toda equations are equivalent to the
equations formulated only in terms of the underlying
finite-dimensional Lie groups and Lie algebras. Actually, partial
cases of such type of equations appeared before, see, for example,
papers \cite{ParShi96, FerGalHolMir97, Mir99} and references
therein, but we demonstrated that any Toda equation of the class
under consideration can be written in terms of finite-dimensional
Lie groups and Lie algebras. To slightly simplify terminology and
make distinction with the Toda equations associated with
finite-dimensional Lie groups, we call the finite-dimensional
version of a Toda equation associated with a loop group of the Lie
group $G$ a loop Toda equation associated with the Lie group $G$.

Here we consider untwisted loop Toda equations associated with
complex general linear group. As was shown in papers
\cite{NirRaz07a, NirRaz07b}, any such equation has the
form\footnote{We denote by $\partial_+$ and $\partial_-$ the partial
derivatives over the standard coordinates $z^+$ and $z^-$ of a
smooth two-dimensional manifold $\calM$, where $\calM$ is either the
Euclidean plane $\bbR^2$ or the complex line $\bbC$; in the latter
case $z^-$ denotes the standard complex coordinate on $\bbC$, and
$z^+$ -- its complex conjugate.}
\begin{equation}
\partial_+ ( \gamma^{-1} \partial_- \gamma )
= [ c_-, \gamma^{-1} c_+ \gamma ],
\label{e:1}
\end{equation}
supplied with the conditions
\begin{equation}
\partial_+ c_- = 0, \qquad
\partial_- c_+ = 0.
\label{e:2}
\end{equation}
Here, $\gamma$ is a mapping of the two-dimensional manifold $\calM$
to the complex general linear group $\rmGL_n(\bbC)$ having a
block-diagonal form
\[
\psset{xunit=1.7em, yunit=1.2em}
\gamma = \left( \raise -1.8\psyunit \hbox{\begin{pspicture}(.6,.6)
(4.5,4.2)
\rput(1,4){$\Gamma_1$}
\rput(2,3){$\Gamma_2$}
\qdisk(2.7,2.3){.7pt} \qdisk(3,2){.7pt} \qdisk(3.3,1.7){.7pt}
\rput(4,1){$\Gamma_p$}
\end{pspicture}} \right),
\]
so that for each $\alpha = 1, \ldots, p$ the mapping $\Gamma_\alpha$
is a mapping of $\calM$ to the Lie group $\rmGL_{n_\alpha}(\bbC)$
with $\sum_{\alpha = 1}^p n_\alpha = n$. Further, $c_+$ and $c_-$
are mappings of $\calM$ to the Lie algebra $\gothgl_n(\bbC)$. The
mapping $c_+$ has the block-matrix structure
\[
\psset{xunit=2.5em, yunit=1.4em}
c_+ = \left( \raise -2.4\psyunit \hbox{\begin{pspicture}(.6,.5)
(5.6,5.3)
\rput(1,5){$0$} \rput(2,4.92){$C_{+1}$}
\rput(2,4){$0$}
\qdisk(2.8,4.2){.7pt} \qdisk(3,4){.7pt} \qdisk(3.2,3.8){.7pt}
\qdisk(2.8,3.2){.7pt} \qdisk(3,3){.7pt} \qdisk(3.2,2.8){.7pt}
\qdisk(3.8,3.2){.7pt} \qdisk(4,3){.7pt} \qdisk(4.2,2.8){.7pt}
\rput(4,2){$0$} \rput(5,1.87){$C_{+(p-1)}$}
\rput(1,.94){$C_{+0}$} \rput(5,1){$0$}
\end{pspicture}} \right),
\]
where for each $\alpha = 1, \ldots, p-1$ the mapping $C_{+\alpha}$ is
a mapping of $\calM$ to the space of $n_\alpha \times n_{\alpha+1}$
complex matrices, and $C_{+0}$ is a mapping of $\calM$ to the space
of $n_p \times n_1$ complex matrices. The mapping $c_-$ has a
similar block-matrix structure,
\[
\psset{xunit=2.5em, yunit=1.4em}
c_- = \left( \raise -2.4\psyunit \hbox{\begin{pspicture}(.5,.5)
(5.5,5.3)
\rput(1,5){$0$} \rput(5,4.92){$C_{-0}$}
\rput(1,4){$C_{-1}$} \rput(2,4){$0$}
\qdisk(1.8,3.2){.7pt} \qdisk(2,3){.7pt} \qdisk(2.2,2.8){.7pt}
\qdisk(2.8,3.2){.7pt} \qdisk(3,3){.7pt} \qdisk(3.2,2.8){.7pt}
\qdisk(2.8,2.2){.7pt} \qdisk(3,2){.7pt} \qdisk(3.2,1.8){.7pt}
\rput(4,1.87){$0$}
\rput(4,.94){$C_{-(p-1)}$} \rput(5,1){$0$}
\end{pspicture}} \right),
\]
where for each $\alpha = 1, \ldots, p-1$ the mapping $C_{-\alpha}$ is
a mapping of $\calM$ to the space of $n_{\alpha+1} \times n_\alpha$
complex matrices, and $C_{-0}$ is a mapping of $\calM$ to the space
of $n_1 \times n_p$ complex matrices. It is assumed that the mappings
$c_+$ and $c_-$ are fixed, and the equation~(\ref{e:1}) is considered
as an equation for the mapping $\gamma$, which can be written explicitly
as a system of equations for the mappings $\Gamma_\alpha$.

It is worth noting that for arbitrary complex classical Lie groups
loop Toda equations belonging to the class under consideration have
the same form (\ref{e:1}) with the same block-matrix structure of
the mappings $\gamma$, $c_+$ and $c_-$, but with some additional
restrictions imposed on the blocks. Note also that the Toda equation
under consideration is Abelian if the mapping $\gamma$ is
effectively a mapping to an Abelian Lie group, otherwise we have a
non-Abelian Toda equation.

In the present paper we consider a particular case of non-Abelian
loop Toda equations associated with the complex general linear group
$\rmGL_n(\bbC)$, where $n_\alpha = n/p  = n_*$ for all
$\alpha = 1,\ldots,p$. Moreover, we assume for simplicity that all
nonzero entries of the block-matrix representation of $c_+$ and $c_-$
are unit $n_* \times n_*$ matrices. In this case the Toda
equation~(\ref{e:1}) can be written as an infinite periodic
system,
\begin{equation}
\partial_+ (\Gamma^{-1}_\alpha \partial_- \Gamma^{}_\alpha)
+ \Gamma^{-1}_\alpha \Gamma^{}_{\alpha+1}
- \Gamma^{-1}_{\alpha-1} \Gamma_{\alpha} = 0,
\label{e:6}
\end{equation}
with $\Gamma_\alpha$ subject to the condition
$\Gamma_{\alpha+p} = \Gamma_\alpha$. This particular case of Toda systems
was introduced in the remarkable paper by Mikhailov \cite{Mik81}.

Here we are interested in explicit solutions of the system (\ref{e:6}),
in particular, in the soliton-like ones in the non-Abelian case when
$n_* > 1$. Soliton solutions of the Abelian loop Toda equations
can be found by various methods. The most known and elaborated among
them are the Hirota's method \cite{Hir04}, successfully applied to many
particular cases of Abelian affine Toda systems \cite{Hol92,
ConFerGomZim93, MacMcG92, AraConFerGomZim93, ZhuCal93, KhaSas96a};
the vertex operators approach of \cite{OliTurUnd93a, OliTurUnd93,
KneOli96} based on a proper specialisation of the Leznov--Saveliev
method \cite{OliSavUnd93}, see also \cite{FerMirGui95, FerMirGui97}
for more details and the relation to the dressing symmetry; and the
formalism of rational dressing developed by Mikhailov \cite{Mik81}
on the basis of a general dressing procedure proposed by Zakharov
and Shabat \cite{ZakSha79}. Actually, in all known cases of Abelian
Toda systems the vertex operators constructions reproduce the same
soliton solutions found by the Hirota's approach. In the paper
\cite{NirRaz08a} we considered Abelian untwisted loop Toda equations
associated with complex general linear groups within the frameworks of
the Hirota's and rational dressing methods and established the explicit
relationships between solutions given by these two approaches. Further,
in the paper \cite{NirRaz08b}, using the rational dressing method, we
have constructed multi-soliton solutions for Abelian twisted loop Toda
systems associated with general linear groups.

There are not many papers dealing with soliton solutions of
non-Abelian loop Toda equations. We would like to mention here paper
\cite{EtiGelRet97a} where a combination of the notion of a
quasi-determinant and the Marchenko lemmas were used to construct
soliton-like solutions of equations (\ref{e:6}). In the paper
\cite{NirRaz08c} we have developed the rational dressing method in
application to non-Abelian untwisted loop Toda equations associated
with complex general linear groups and found certain multi-soliton
solutions. Here we restrict our attention to equations (\ref{e:6})
and show that the rational dressing method in this most symmetric
case allows one to construct new soliton-like solutions which can be
presented in a form of a direct matrix generalization of the
Hirota's soliton solutions well-known for the case of Abelian loop
Toda systems.

The main idea of this paper, as of the paper \cite{NirRaz08c}, is to
demonstrate the power of the rational dressing formalism appropriately
developed for explicitly constructing solutions to the non-Abelian loop
Toda equations. It is a distinctive feature of this method that it allows
for such remarkable generalizations to the non-Abelian case, where other
well-known methods fail to work. Among the solutions to be presented in
what follows, we single out a class of soliton-like ones thus justifying
the title of our paper. Here, by an $n$-soliton, or soliton-like, solution
we mean a solution depending on $n$ linear combinations of independent
variables and having an appropriate number of characteristic parameters.

\section{Rational dressing}

We see that the constant matrices $c_-$ and $c_+$ commute.
Hence, it is obvious that
\begin{equation}
\gamma = I_n,
\label{e:7}
\end{equation}
where $I_n$ is the $n \times n$ unit matrix, is a solution to the
Toda equation (\ref{e:1}). For the formalism of rational dressing
it is crucial that the Toda equation (\ref{e:1}), (\ref{e:2}) can
be represented as the zero-curvature condition for a flat connection
in a trivial principal fiber bundle, satisfying the grading and
gauge-fixing conditions, see, for example, the books
\cite{LezSav92, RazSav97}. Actually, having such a connection we
have a solution to the Toda equations (\ref{e:1}). In our case,
the corresponding base manifold $\calM$ is either the Euclidean
plane $\bbR^2$ or the complex manifold $\bbC$, and the fiber
coincides with the untwisted loop group $\calL_{a,p}(\rmGL_n(\bbC))$,
where $a$ denotes the inner automorphism of $\rmGL_n(\bbC)$ of order
$p$ acting on an element $g \in \rmGL_n(\bbC)$ in accordance with
the equality
\[
a(g) = h \: g \: h^{-1}.
\]
Here $h$ is a block-diagonal matrix defined by the relation
\begin{equation}
h^{}_{\alpha\beta}
= \epsilon_p^{p - \alpha + 1} I_{n_*} \delta^{}_{\alpha\beta},
\qquad \alpha, \beta = 1, \ldots, p, \label{e:h}
\end{equation}
where $\epsilon_p = \rme^{2\pi\rmi/p}$ is the $p$th principal root of
unity.

Using the exponential law \cite{KriMic91,KriMic97}, it is convenient
to identify the mapping generating a flat connection under consideration
with a smooth mapping of $\calM \times S^1$ to $\rmGL_n(\bbC)$ and the
connection components with smooth mappings of $\calM \times S^1$ to
$\gothgl_n(\bbC)$. Below we think of the circle $S^1$ as consisting of
complex numbers of modulus one.

Denote the mapping generating the connection corresponding to the
solution (\ref{e:7}) by $\varphi$. For the case under consideration,
the rational dressing method consists in finding a mapping $\psi$ of
$\calM \times S^1$ to $\rmGL_n(\bbC)$, such that the grading and
gauge-fixing conditions for the components of the flat connection
generated by the mapping $\varphi \: \psi$ are satisfied. We assume
that the analytic extension of the mapping $\psi$ from $S^1$ to the
whole Riemann sphere is a rational mapping given by the expression
\begin{equation}
\psi = \left( I_n + \sum_{i=1}^r \sum_{k=1}^p \frac{\lambda}
{\lambda - \epsilon_p^k \mu_i} h^k P_i \: h^{-k} \right) \psi^{}_0.
\label{e:8}
\end{equation}
Here $\lambda$ is the standard coordinate in $\bbC$, $\psi_0$ is a
mapping of $\calM$ to the Lie subgroup of $\rmGL_n(\bbC)$ formed by
the elements $g \in \rmGL_n(\bbC)$ subject to the equality $h \: g
\: h^{-1} = g$, and $P_i$ are some smooth mappings of $\calM$ to the
algebra $\Mat_n(\bbC)$ of $n \times n$ complex matrices. It is also
assumed here that $\mu_i \ne 0$, $\mu_i^p \ne \mu_j^p$ for all $i
\ne j$. Expression (\ref{e:8}) is obtained from an initial rational
mapping by averaging over the action of the inner automorphism $a$,
where it was used that $a^p = \id_{\rmGL(\bbC)}$. Further, we
suppose that the analytic extension of the corresponding inverse
mapping to the whole Riemann sphere has a similar structure,
\[
\psi^{-1} = \psi^{-1}_0 \left( I_n
+ \sum_{i = 1}^r \sum_{k=1}^p \frac{\lambda}
{\lambda - \epsilon_p^k \nu_i} \: h^k \: Q_i \: h^{-k} \right),
\]
with the pole positions satisfying the conditions $\nu_i \ne 0$,
$\nu_i^p \ne \nu_j^p$ for all $i \ne j$, and additionally $\nu_i^p \ne
\mu_j^p$ for any $i$ and $j$. Note that the mappings $\psi(\lambda)$
and $\psi^{-1}(\lambda)$ are regular at the points $\lambda = 0$ and
$\lambda = \infty$.

By definition, the equality
\[
\psi^{-1} \psi = I_n
\]
is valid on $S^1$. Since $\psi$ and $\psi^{-1}$ are rational mappings,
this equality is valid on the whole Riemann sphere. Therefore, the
residues of $\psi^{-1} \psi$ at the points $\mu_i$ and $\nu_i$ must
vanish. This leads to certain relations to be satisfied by the mappings
$P_i$ and $Q_i$,
\begin{gather}
Q_i \left( I_n + \sum_{j = 1}^r \sum_{k=1}^p \frac{\nu_i}{\nu_i -
\epsilon_p^k \: \mu_j} \: h^k \: P_j \: h^{-k} \right) = 0,
\label{e:res1_n} \\
\left( I_n + \sum_{j = 1}^r \sum_{k=1}^p \frac{\mu_i}{\mu_i -
\epsilon_p^k \: \nu_j} \: h^k \: Q_j \: h^{-k} \right) P_i = 0.
\label{e:res1_m}
\end{gather}
Further, for the components of the flat connection generated
by the mapping $\varphi \: \psi$ we find the expressions
\begin{gather*}
\omega_- = \psi^{-1} \partial_- \psi
+ \lambda^{-1} \psi^{-1} c_- \psi,  \\
\omega_+ = \psi^{-1} \partial_+ \psi
+ \lambda \psi^{-1} c_+ \psi.
\end{gather*}
We see that $\omega_-$ is a rational mapping having simple poles
at $\mu_i$, $\nu_i$ and zero. Similarly, $\omega_+$ is a rational
mapping having simple poles at $\mu_i$, $\nu_i$ and infinity. We
need a connection satisfying the grading and gauge-fixing conditions.
The grading condition in our case means that for each point of $\calM$
the component $\omega_-(\lambda)$ is rational and has the only simple
pole at zero, and the component $\omega_+(\lambda)$ is rational and has
the only simple pole at infinity. Therefore, we require that the residues
of $\omega_-$ and $\omega_+$ at the points $\mu_i$ and $\nu_i$ must vanish.
And this requirement imposes additional conditions on the mappings $P_i$
and $Q_i$, that are
\begin{gather}
(\partial_- Q_i - \nu_i^{-1} Q_i \: c_-) \left( I_n + \sum_{j = 1}^r
\sum_{k=1}^p \frac{\nu_i}{\nu_i - \epsilon_p^k \: \mu_j} \: h^k \: P_j
\: h^{-k}
\right) = 0,
\label{e:res2_n} \\
(\partial_+ Q_i - \nu_i \: Q_i \: c_+) \left( I_n + \sum_{j = 1}^r
\sum_{k=1}^p \frac{\nu_i}{\nu_i - \epsilon_p^k \: \mu_j} \: h^k \: P_j
\: h^{-k}
\right) = 0
\label{e:res3_n}
\end{gather}
for the residues at the points $\nu_i$, and also
\begin{gather}
\left( I_n + \sum_{j = 1}^r \sum_{k=1}^p \frac{\mu_i}{\mu_i -
\epsilon_p^k \: \nu_j} \: h^k \: Q_j \: h^{-k} \right)
(\partial_- P_i + \mu_i^{-1} c_- \: P_i) = 0,
\label{e:res2_m} \\
\left( I_n + \sum_{j = 1}^r \sum_{k=1}^p \frac{\mu_i}{\mu_i -
\epsilon_p^k \: \nu_j} \: h^k \: Q_j \: h^{-k} \right)
(\partial_+ P_i + \mu_i \: c_+ \: P_i) = 0
\label{e:res3_m}
\end{gather}
for the residues at the points $\mu_i$. Having these and the above
mentioned relations fulfilled by $P_i$ and $Q_i$, and besides assuming
that $\psi_0 = I_n$ that resolves the gauge-fixing constraint
$\omega_{+0} = 0$, where we put $\omega_{+0} = \omega_+(0)$, we can
see that the mapping $\gamma = \psi(\infty)$, with fixed $2r$ complex
numbers $\mu_i$, $\nu_i$, satisfies the Toda equation (\ref{e:1}).
In the block-matrix form we have the expressions
\begin{equation}
\gamma^{}_{\alpha\beta} = \delta^{}_{\alpha\beta} \left(
I_{n_*} + p \sum^r_{i = 1} (P_i)^{}_{\alpha\alpha} \right), \qquad
\gamma^{-1}_{\alpha\beta} = \delta^{}_{\alpha\beta} \left(
I_{n_*} + p \sum^r_{i = 1} (Q_i)^{}_{\alpha\alpha} \right),
\label{e:13}
\end{equation}
where $(P_i)^{}_{\alpha\beta}$ and $(Q_i)^{}_{\alpha\beta}$ are
$n_* \times n_*$ complex matrices satisfying certain conditions
(\ref{e:res1_n})--(\ref{e:res3_m}). These conditions ensuring the
vanishing of the residues of $\psi^{-1} \psi$, $\omega_-$ and
$\omega_+$ at the points $\nu_i$, $\mu_i$ can be non-trivially
fulfilled, see also the papers \cite{NirRaz08a,NirRaz08c}.

To make the solution (\ref{e:13}) explicit we should specify the
matrix-valued functions $P_i$ and $Q_i$. We first note that, if we
suppose that the functions $P_i$ and $Q_i$ take values in the space
of the matrices of maximum rank, then we come to the trivial
solution (\ref{e:7}). Hence, we assume that the values of the
functions $P_i$ and $Q_i$ are not matrices of maximum rank. The case
given by matrices of rank one was elaborated in paper
\cite{NirRaz08c}. Now we consider another interesting case, where
the functions $P_i$ and $Q_i$ take values in the space of $n \times
n$ matrices of rank $n_*$. Such functions can be represented
as\footnote{Hereafter, the superscript $t$ stands for the usual
matrix transposition.}
\begin{equation}
P_i = u^{}_i {}^{t\!} w^{}_i, \qquad Q_i = x^{}_i {}^{t\!} y^{}_i,
\label{e:14}
\end{equation}
where $u$, $w$, $x$ and $y$ are functions on $\calM$ taking values
in the space of $n \times n_*$ complex matrices of rank $n_*$.
The used $\bbZ$-gradation suggests the most convenient
block-matrix representation for the matrices of the form
(\ref{e:14}) as
\[
(P_i)_{\alpha\beta} = u_{i,\alpha} \: {}^{t\!}w_{i,\beta},
\qquad
(Q_i)_{\alpha\beta} = x_{i,\alpha} \: {}^{t\!}y_{i,\beta},
\]
where the standard matrix multiplication of the $n_* \times n_*$
matrix-valued functions $u_{i,\alpha}$, $x_{i,\alpha}$ by the $n_*
\times n_*$ matrix-valued functions ${}^{t\!}w_{i,\beta}$,
${}^{t\!}y_{i,\beta}$ is implied. Considering the rank of the
matrices $P_i$ and $Q_i$ not to be equal to $1$, we are looking for
a novel generalization of the Abelian loop Toda soliton
constructions \cite{NirRaz08a} that would be different from those
given in paper \cite{NirRaz08c}.

Then, within the formalism of rational dressing we see that the functions
$w$ and $x$ can be expressed via the functions $y$ and $u$, and we find
the expressions
\[
(P_i)_{\alpha\beta} = - \frac{1}{p} u_{i,\alpha} \sum_{j=1}^{r}
(R^{-1}_\beta)_{i j} \: {}^{t\!}y_{j,\beta}, \qquad
(Q_i)_{\alpha\beta} = \frac{1}{p} \sum_{j=1}^{r} u_{j,\alpha} \:
\frac{1}{\mu_j} \: (R^{-1}_{\alpha+1})_{j i} \: \nu_i \:
{}^{t\!}y_{i,\beta},
\]
with ${n_* r} \times {n_* r}$ matrix-valued functions $R_\alpha$ defined
through its ${n_*} \times {n_*}$ blocks as
\[
(R_\alpha)_{i j} = \frac{1}{\nu_i^p - \mu_j^p}
\sum_{\beta=1}^p \nu_i^{p - |\beta-\alpha|_p} \mu_j^{|\beta-\alpha|_p}
\: {}^{t\!}y_{i,\beta} \: u_{j,\beta},
\]
where $|x|_p$ denotes the residue of division of $x$ by $p$. It is
important to note here that, unlike the Abelian and non-Abelian
cases considered in papers \cite{NirRaz08a, NirRaz08b, NirRaz08c},
for any $i, j$ the block $(R_\alpha)_{i j}$ is now an $n_* \times
n_*$ matrix-valued function, that adds to the explicitly indicated
summations over the pole indices respective matrix multiplications.

It is convenient to use quantities defined as
$\wt u_{i,\alpha} = u_{i,\alpha} \mu^\alpha_i$,
$\wt y_{i,\alpha} = y_{i,\alpha} \nu^{-\alpha}_i$ and
$(\wt R_\alpha)_{i j} = \nu^{-\alpha}_i (R_\alpha)_{i j} \: \mu^\alpha_j$.
For the matrices $\wt R_\alpha$ we have explicitly the
relation
\[
(\wt R_\alpha)_{i j} = \frac{1}{\nu_i^p - \mu_j^p}
\left( \mu_j^p \sum_{\beta=1}^{\alpha-1}  {}^{t\!} \wt y_{i,\beta}
\: \wt u_{j,\beta}
+ \nu_i^p \sum_{\beta=\alpha}^p {}^{t\!} \wt y_{i,\beta}
\: \wt u_{j,\beta} \right).
\]
Then, in terms of these quantities, the $n_* \times n_*$
matrix-valued functions $\Gamma_\alpha$ can be written as
\[
\Gamma_\alpha = I_{n_\alpha} - \sum_{i,j=1}^{r}
\wt u_{i,\alpha} \: (\wt R^{-1}_\alpha)_{i j}
\: {}^{t\!} \wt y_{j,\alpha}.
\]
Similarly, for the corresponding inverse mappings
we obtain the expression
\[
\Gamma^{-1}_\alpha = I_{n_\alpha} + \sum^{r}_{i,j=1}
\wt u^{}_{i,\alpha}
(\wt R^{-1}_{\alpha+1})_{i j} \: {}^{t\!} \wt y_{j,\alpha},
\]
which can be useful for verifying the Toda equations.

To finally satisfy the conditions imposed earlier on the
matrix-valued functions $P_i$ and $Q_i$, we also demand
the validity of the equations
\begin{gather}
\partial_- u_i = - \mu_i^{-1} \: c_- \: u_i,
\qquad
\partial_+ u_i = - \mu_i^{} \: c_+ \: u_i,
\label{e:21} \\
\partial_- y_i = \nu_i^{-1} \: {}^{t\!}c_- \: y_i,
\qquad
\partial_+ y_i = \nu_i^{} \: {}^{t\!}c_+ \: y_i,
\label{e:22}
\end{gather}
that are sufficient to fulfill relations (\ref{e:res2_n}),
(\ref{e:res3_n}) and (\ref{e:res2_m}), (\ref{e:res3_m}). Using the
explicit forms of the matrices $c_\pm$, we write down the general
solutions to (\ref{e:21}), (\ref{e:22}) as\footnote{It should be
instructive to confer these expressions with those ones derived in
paper \cite{NirRaz08c} for the case of general $\bbZ$-gradations of
inner type.}
\begin{gather}
u_{i,\beta} = \sum^p_{\alpha=1} \epsilon^{\beta\alpha}_{p}
\exp \left( -\mu^{-1}_i \: \epsilon_p^{-\alpha} \: z^-
- \mu_i \: \epsilon_p^\alpha \: z^+ \right) c_{i,\alpha},
\label{e:23} \\
y_{i,\beta} = \sum^p_{\alpha=1} \epsilon^{\beta\alpha}_{p}
\exp \left( \nu^{-1}_i \: \epsilon_p^{\alpha} \: z^-
+ \nu_i \: \epsilon_p^{-\alpha} \: z^+ \right) d_{i,\alpha},
\label{e:24}
\end{gather}
where $c_{i,\alpha}$ and $d_{i,\alpha}$ are $n_* \times n_*$ complex
matrices meaning the initial-value data for equations (\ref{e:21}),
(\ref{e:22}). With these solutions we immediately obtain for the
blocks of the matrix-valued functions $\wt R_\alpha$ the following
expression:
\[
(\wt R_\alpha)_{i j} = \sum^p_{\beta,\delta=1} \rme^{Z_{-\beta}(\nu_i)
- Z_\delta(\mu_j)}
\frac{\epsilon_p^{\alpha(\beta+\delta)}}{1-\mu_j\nu^{-1}_i
\epsilon_p^{\beta+\delta}} \: \nu^{-\alpha}_i
({}^{t\!}d_{i,\beta} \: c_{j,\delta}) \: \mu^\alpha_j,
\]
where we have introduced the notation
$Z_\alpha(\mu_i) = \mu^{-1}_i \epsilon^{-\alpha}_p z^-
+ \mu^{}_i \: \epsilon^{\alpha}_p \: z^+$.

\section{Soliton-like solutions}

To construct solutions making sense as $r$-solitons, that is, by the
definition we use here, solutions depending on $r$ linear combinations
of independent variables $z^+$ and $z^-$, we assume that for each value
of the index $i = 1, \ldots, r$ the initial-value data of the Toda system
under consideration are such that matrix-valued coefficients $c_{i,\alpha}$
are different from zero for only one value of $\alpha$, which we denote
by $I_i$, and that the matrix-valued coefficients $d_{i,\alpha}$ are
different from zero for only two values of $\alpha$, which we denote
by $J_i$ and $K_i$. We also use for such non-vanishing initial-data
$n_* \times n_*$ matrices the notation $d_{J_i} = d_{i,J_i}$,
$d_{K_i} = d_{i,K_i}$ and $c_{I_i} = c_{i,I_i}$. For the $n_* \times n_*$
blocks of the matrix-valued functions $\wt u_i$ and $\wt y_i$ this
assumption gives
\begin{align}
& \wt u^{}_{i,\alpha} = \mu_i^{\alpha} \: \epsilon_p^{\alpha I_i}
\: \rme^{-Z_{I_i}(\mu_i)} \: c_{I_i},
\label{e:26} \\[.5em]
& {}^{t\!}\wt y_{i,\alpha} = \nu_i^{-\alpha} \: \epsilon_p^{\alpha J_i}
\: \rme^{Z_{-J_i}(\nu_i)} \: {}^{t\!}d_{J_i}
+ \nu_i^{-\alpha} \: \epsilon_p^{\alpha K_i} \: \rme^{Z_{-K_i}(\nu_i)}
\: {}^{t\!}d_{K_i}.
\label{e:27}
\end{align}
With these relations, we can write the expression for the mappings
$\Gamma_\alpha$ in the form
\begin{equation}
\Gamma_\alpha = I_{n_\alpha} - \sum^{r}_{i,j=1} c_{I_i}
(\wt R^{\prime -1}_\alpha)^{}_{i j}
({}^{t\!}d_{J_j} + E_{\alpha,j} \:{}^{t\!}d_{K_j} ),
\label{e:28}
\end{equation}
where we have used the notation
\begin{equation}
(\wt R'_\alpha)^{}_{i j}
= \wt D_{i j}(J) + E_{\alpha,i} \: \wt D_{i j}(K), \qquad
E_{\alpha,i} = \epsilon_p^{\alpha \rho_i} \rme^{Z_i(\zeta)},
\label{e:29}
\end{equation}
with the dependence on the variables $z^+$ and $z^-$ given through
the functions
\[
Z_i(\zeta) = \kappa_{\rho_i} (\zeta_i^{-1} z^- + \zeta_i^{} z^+),
\]
and the convenient parameters $\rho_i = K_i - J_i$,
$\zeta_i = -\rmi \nu_i \epsilon_p^{-(K_i + J_i)/2}$,
$\kappa_{\rho_i} = 2\sin(\pi\rho/p)$, and besides
\begin{equation}
\wt D_{i j}(A) = \frac{{}^{t\!}d_{A_i} \: c_{I_j}}
{1 -\nu^{-1}_i \mu^{}_j \epsilon_p^{A_i + I_j}},
\qquad i,j = 1,\ldots,r,
\label{e:30}
\end{equation}
for the $n_* \times n_*$ blocks of the $n_* r \times n_* r$
matrices $\wt D(A)$, $A = J, K$. Similar $r \times r$
matrices were introduced already in our previous paper
\cite{NirRaz08c} for the rank-$1$ case, however now, in
the case of rank-$n_*$, we have a different situation,
such that $\wt D_{ij}$ itself is a complex $n_* \times n_*$
matrix for each $i$ and $j$.

We assume that the $n_* \times n_*$ matrices $c_{I_i}$ are
non-degenerate. Let us multiply $\Gamma_\alpha$ in (\ref{e:28})
by $c_{I_\ell}$ from the right hand side and write
\begin{equation}
\Gamma_\alpha \: c_{I_\ell}= \sum^{r}_{i,j=1} c_{I_i}
(\wt R^{\prime -1}_{\alpha})^{}_{i j}
[(\wt R^{\prime}_{\alpha})^{}_{j \ell} -
({}^{t\!}d_{J_j} + E_{\alpha,j} \:{}^{t\!}d_{K_j}) c_{I_\ell}].
\label{e:31}
\end{equation}
It is not difficult to see that
\[
(\wt R^{\prime}_{\alpha})^{}_{j \ell} -
({}^{t\!}d_{J_j} + E_{\alpha,j} \:{}^{t\!}d_{K_j}) c_{I_\ell}
= \nu^{-1}_j \: \epsilon_p^{J_j} (\wt R^{\prime}_{\alpha+1})^{}_{j \ell}
\mu^{}_\ell \: \epsilon_p^{I_\ell}.
\]
Now, multiplying (\ref{e:31}) from the right hand side by
the inverse matrix $c^{-1}_{I_{\ell}}$ and summing up over
$\ell = 1,\ldots,r$, we obtain the following expression:
\[
\Gamma_\alpha = \frac{1}{r} \sum^{r}_{i,j,k=1} c_{I_i}
(\wt R^{\prime -1}_{\alpha})^{}_{i j}
\nu^{-1}_j \: \epsilon_p^{J_j} (\wt R^{\prime}_{\alpha+1})^{}_{j k}
\mu^{}_k \: \epsilon_p^{I_k} \: c^{-1}_{I_k}.
\]
To give the soliton solutions a final form, it is also convenient to
make use of simplest symmetries of the Toda equation. It is clear
from relations (\ref{e:6}) that the transformations
\begin{equation}
\Gamma_\alpha \to \xi \Gamma_\alpha, \qquad
\Gamma_\alpha \to x^{-1} \Gamma_\alpha x^{}
\label{e:34}
\end{equation}
for a nonzero constant $\xi$ and a non-singular constant
$n_* \times n_*$ matrix $x$ are symmetry transformations
of the Toda system under consideration.

In particular, using the symmetry transformations (\ref{e:34})
with $\xi = \mu^{-1} \nu \epsilon_p^{-(I+J)}$ and $x = c_I$, we
can write the one-soliton solution as
\begin{equation}
\Gamma_\alpha = \wt R^{\prime -1}_{\alpha} \:
\wt R^{\prime}_{\alpha+1},
\label{e:35}
\end{equation}
where the matrices $\wt R^{\prime}_{\alpha}$ are explicitly given
by the relations (\ref{e:29}), (\ref{e:30}). It reproduces exactly
the one-soliton solution constructed in the paper \cite{EtiGelRet97a}
by means of a different approach. The solution (\ref{e:35}) can also
be written in a convenient form
\[
\Gamma_\alpha = T^{-1}_\alpha T^{}_{\alpha+1},
\]
where $T_\alpha = I_{n_\alpha} + E_\alpha H$ and
$H = \wt D(K) \wt D^{-1}(J)$.

Also the multi-soliton solutions, $r \ge 2$, can be written
in a compact form,
\begin{equation}
\Gamma_\alpha = {}^{T\!}c_I (\wt R^{\prime -1}_{\alpha}) \:
\calN_J^{-1} \: (\wt R^{\prime}_{\alpha+1}) \: \calM^{}_I \:
c^{-1}_I.
\label{e:36}
\end{equation}
Here we use the notation ${}^{T\!}c_I$ and $c^{-1}_I$ for
$n_* \times n_*r$ and $n_*r \times n_*$ matrices, respectively,
defined as ${}^{T\!}c_I = (c_{I_1} \: \ldots \: c_{I_r})$
and
${}^{t\!}c^{-1}_I = ({}^{t\!}c^{-1}_{I_1} \: \ldots \:{}^{t\!}c^{-1}_{I_r})$,
and the notation $\calN_J^{-1}$ and $\calM_I$ for block-diagonal
$n_* r \times n_* r$ matrices defined as
\[
\calN_J^{-1} = \left( \begin{array}{ccc}
\nu_1^{-1} \epsilon_p^{J_1} I_{n_*} & & \\
                                  & \ddots & \\
& & \nu_r^{-1} \epsilon_p^{J_r} I_{n_*}
                      \end{array}
\right),
\qquad
\calM_I^{} = \left( \begin{array}{ccc}
\mu_1^{} \epsilon_p^{I_1} I_{n_*} & & \\
                                  & \ddots & \\
& & \mu_r^{} \epsilon_p^{I_r} I_{n_*}
                      \end{array}
\right).
\]
Putting $n_* = 1$ we can recover the Abelian case analyzed in paper
\cite{NirRaz08a}. The solutions (\ref{e:35}) and (\ref{e:36}) may be
regarded as a novel non-Abelian generalization, complementary to
that of paper \cite{NirRaz08c}, of the Hirota's soliton
constructions \cite{Hir04} successfully used for the Abelian loop
Toda systems. Recall that the $\tau$-functions of the Abelian
construction of Hirota, the $\tau_\alpha$, were generalized by the
set of matrix-valued functions $\widetilde{T}^X_\alpha$ and ordinary
functions $\widetilde{T}^{}_{\alpha}$ in the non-Abelian case
\cite{NirRaz08c}, while now, in a higher-rank case, we have the set
of matrix-valued functions $\wt R^{\prime}_{\alpha+1} \: \calM^{}_I
\: c^{-1}_I$ and ${}^{T\!}c_I \wt R^{\prime -1}_{\alpha} \:
\calN_J^{-1}$, taken always in a respective combination, instead of
the functions $\tau_\alpha$.

To obtain more solutions to these equations, not necessarily
soliton-like, one should keep nonzero more initial-value data
$c_{i,\alpha}$ and $d_{i,\alpha}$ entering (\ref{e:23}), (\ref{e:24}),
that leads to more general expressions for $\wt u^{}_{i,\alpha}$ and
$\wt y^{}_{i,\alpha}$ than (\ref{e:26}) and (\ref{e:27}).

\section{Discussion}

It should be rather illuminative to reproduce our results along
the lines of any other approaches. One such a possibility would
be, probably, to try the general dressing procedure \cite{BabBer92,
FerMirGui95, FerMirGui97}. Here, after the dressing transformations
are completed, the problem is reduced to certain spectral problems
for the matrices $c_+$ and $c_-$. Thus, using specific vertex
operators $V_i$ related to an appropriate basis, one could
bring the generalized $\tau$-functions to the form
\[
\prod^r_{i = 1} \left(I_{n_\alpha} + E_{\alpha,i} V_i
+ \ldots \right),
\]
where the vertex operators are supposed to obey certain nilpotency
conditions \cite{FerMirGui95, FerMirGui97}. As we mentioned at the
beginning, the resulting expressions can be then directly compared
to what one finds in the case of Abelian Toda systems. However, it
is not clear yet how to provide the corresponding statement for
the non-Abelian case, while preserving the block-matrix structure
suggested by the $\bbZ$-gradation.

The next step, most naturally following our constructions, would be
a thorough investigation of the physical content of the solutions.
In this way, one should describe standard properties defining the
solitons, for example, in the spirit of paper \cite{Hol92}. Note
that such a program, in general, must be based on the specification
of real forms of the loop Toda equations under consideration, so
that the solutions making sense as `physical solitons' could be
found in the present ones by certain reductions. Here, such
reductions impose certain conditions on the characteristic
parameters entering the explicit forms of the soliton-like
solutions. In particular, it can be shown that to compact real forms
specified by the conditions $\Gamma^\dagger_\alpha =
\Gamma^{-1}_\alpha$, where dagger means the Hermitian conjugation,
actually correspond such `physical solitons', and there are no such
solutions in the case of non-compact real forms corresponding to the
condition $\Gamma^*_\alpha = \Gamma^{}_\alpha$, where star denotes
the usual complex conjugation.

A few comments are in order. To have a solution making sense as
a soliton, we must require that $\mu^*_i = \nu_i$. Putting $r=1$
we see that it should be valid $ H^\dagger \equiv (H' \exp\delta)^\dagger
= \epsilon^\rho_p H' \exp\delta$, where we have used the notation
$\exp\delta = (1 - \mu\nu^{-1}\epsilon_p^{I+J})
/(1 - \mu\nu^{-1}\epsilon_p^{I+K})$. Note that the function
$Z(\zeta)$ should be real, and so, if $z^- = x - \mathrm{i} t$,
$z^+ = x + \mathrm{i} t$, this function can be written in a familiar
form, as $2\kappa_\rho(x - vt)/\sqrt{1 + v^2}$, where $v$ is the ratio
of the imaginary and real parts of the parameter $\zeta$, $|\zeta|^2 = 1$,
and if $z^- = x - t$, $z^+ = x + t$, as $2\kappa_\rho(x + vt)/\sqrt{1 - v^2}$,
where $v = (\zeta - 1/\zeta) / (\zeta + 1/\zeta)$ for a real $\zeta$.
In the Abelian limit, when $n_* = 1$, the matrix $H'$ is just a complex
number, and it can be lifted up to $\exp\delta$, and then the above
relation, that is a restriction on the initial-value data, fixes the
imaginary part of total $\delta$ to be $-\pi\rho/p$.

We would like to refer to the paper \cite{NirRaz07b}, where the
above real forms of loop Toda equations were obtained for the case
of $p=2$, and to the paper \cite{ParShi96}, where basic physical
properties of one-soliton and two-soliton (soliton -- anti-soliton
and breather) solutions were investigated for a matrix
generalization of the sine-Gordon equation based on the coset
$\mathrm{SU}_2 \times \mathrm{SU_2} / \mathrm{SU_2}$. In the Abelian
case the physical properties of loop Toda solitons, including
masses, topological charges, scattering processes, were described in
paper \cite{Hol92}. We will address these issues about our
non-Abelian constructions in forthcoming publications.

\vskip3mm
{\bf Acknowledgments}

We express our gratitude to Profs. H. Boos, F. G\"ohmann and
A. Kl\"umper for hospitality at the University of Wuppertal
and interesting discussions. This work was supported in part
by the by the joint DFG--RFBR grant \#08--01--91953.

\end{document}